
\input epsf
\font \tbfontt                = cmbx10 scaled\magstep1
\font \tafontt                = cmbx10 scaled\magstep2
\font \tbfontss               = cmbx5  scaled\magstep1
\font \tafontss               = cmbx5  scaled\magstep2
\font \sixbf                  = cmbx6
\font \tbfonts                = cmbx7  scaled\magstep1
\font \tafonts                = cmbx7  scaled\magstep2
\font \ninebf                 = cmbx9
\font \tasys                  = cmex10 scaled\magstep1

\font \sixi                   = cmmi6
\font \ninei                  = cmmi9
\font \tams                   = cmmib10
\font \tbmss                  = cmmib10 scaled 600
\font \tamss                  = cmmib10 scaled 700
\font \tbms                   = cmmib10 scaled 833
\font \tbmt                   = cmmib10 scaled\magstep1
\font \tamt                   = cmmib10 scaled\magstep2
\font \sixrm                  = cmr6
\font \ninerm                 = cmr9
\font \ninesl                 = cmsl9
\font \tensans                = cmss10
\font \fivesans               = cmss10 at 5pt
\font \sixsans                = cmss10 at 6pt
\font \sevensans              = cmss10 at 7pt
\font \ninesans               = cmss10 at 9pt
\font \tbst                   = cmsy10 scaled\magstep1
\font \tast                   = cmsy10 scaled\magstep2
\font \tbsss                  = cmsy5  scaled\magstep1
\font \tasss                  = cmsy5  scaled\magstep2
\font \sixsy                  = cmsy6
\font \tbss                   = cmsy7  scaled\magstep1
\font \tass                   = cmsy7  scaled\magstep2
\font \ninesy                 = cmsy9

\font \nineit                 = cmti9
\font \ninett                 = cmtt9
\magnification=\magstep0
\hsize=11.7truecm
\vsize=19.0truecm
\hfuzz=2pt
\tolerance=500
\abovedisplayskip=3 mm plus6pt minus 4pt
\belowdisplayskip=3 mm plus6pt minus 4pt
\abovedisplayshortskip=0mm plus6pt minus 2pt
\belowdisplayshortskip=2 mm plus4pt minus 4pt
\predisplaypenalty=0
\clubpenalty=10000
\widowpenalty=10000
\frenchspacing
\newdimen\oldparindent\oldparindent=1.5em
\parindent=1.5em

\def\arcmin{\hbox{$^\prime$}}

\def\utw{\smash{\rlap{\lower5pt\hbox{$\sim$}}}}
\def\udtw{\smash{\rlap{\lower6pt\hbox{$\approx$}}}}



\def\bbbc{{\mathchoice {\setbox0=\hbox{$\displaystyle\rm C$}\hbox{\hbox
to0pt{\kern0.4\wd0\vrule height0.9\ht0\hss}\box0}}
{\setbox0=\hbox{$\textstyle\rm C$}\hbox{\hbox
to0pt{\kern0.4\wd0\vrule height0.9\ht0\hss}\box0}}
{\setbox0=\hbox{$\scriptstyle\rm C$}\hbox{\hbox
to0pt{\kern0.4\wd0\vrule height0.9\ht0\hss}\box0}}
{\setbox0=\hbox{$\scriptscriptstyle\rm C$}\hbox{\hbox
to0pt{\kern0.4\wd0\vrule height0.9\ht0\hss}\box0}}}}
\def\bbbq{{\mathchoice {\setbox0=\hbox{$\displaystyle\rm Q$}\hbox{\raise
0.15\ht0\hbox to0pt{\kern0.4\wd0\vrule height0.8\ht0\hss}\box0}}
{\setbox0=\hbox{$\textstyle\rm Q$}\hbox{\raise
0.15\ht0\hbox to0pt{\kern0.4\wd0\vrule height0.8\ht0\hss}\box0}}
{\setbox0=\hbox{$\scriptstyle\rm Q$}\hbox{\raise
0.15\ht0\hbox to0pt{\kern0.4\wd0\vrule height0.7\ht0\hss}\box0}}
{\setbox0=\hbox{$\scriptscriptstyle\rm Q$}\hbox{\raise
0.15\ht0\hbox to0pt{\kern0.4\wd0\vrule height0.7\ht0\hss}\box0}}}}
\def\bbbt{{\mathchoice {\setbox0=\hbox{$\displaystyle\rm
T$}\hbox{\hbox to0pt{\kern0.3\wd0\vrule height0.9\ht0\hss}\box0}}
{\setbox0=\hbox{$\textstyle\rm T$}\hbox{\hbox
to0pt{\kern0.3\wd0\vrule height0.9\ht0\hss}\box0}}
{\setbox0=\hbox{$\scriptstyle\rm T$}\hbox{\hbox
to0pt{\kern0.3\wd0\vrule height0.9\ht0\hss}\box0}}
{\setbox0=\hbox{$\scriptscriptstyle\rm T$}\hbox{\hbox
to0pt{\kern0.3\wd0\vrule height0.9\ht0\hss}\box0}}}}
\def\bbbs{{\mathchoice
{\setbox0=\hbox{$\displaystyle     \rm S$}\hbox{\raise0.5\ht0\hbox
to0pt{\kern0.35\wd0\vrule height0.45\ht0\hss}\hbox
to0pt{\kern0.55\wd0\vrule height0.5\ht0\hss}\box0}}
{\setbox0=\hbox{$\textstyle        \rm S$}\hbox{\raise0.5\ht0\hbox
to0pt{\kern0.35\wd0\vrule height0.45\ht0\hss}\hbox
to0pt{\kern0.55\wd0\vrule height0.5\ht0\hss}\box0}}
{\setbox0=\hbox{$\scriptstyle      \rm S$}\hbox{\raise0.5\ht0\hbox
to0pt{\kern0.35\wd0\vrule height0.45\ht0\hss}\raise0.05\ht0\hbox
to0pt{\kern0.5\wd0\vrule height0.45\ht0\hss}\box0}}
{\setbox0=\hbox{$\scriptscriptstyle\rm S$}\hbox{\raise0.5\ht0\hbox
to0pt{\kern0.4\wd0\vrule height0.45\ht0\hss}\raise0.05\ht0\hbox
to0pt{\kern0.55\wd0\vrule height0.45\ht0\hss}\box0}}}}
\def\bbbz{{\mathchoice {\hbox{$\sans\textstyle Z\kern-0.4em Z$}}
{\hbox{$\sans\textstyle Z\kern-0.4em Z$}}
{\hbox{$\sans\scriptstyle Z\kern-0.3em Z$}}
{\hbox{$\sans\scriptscriptstyle Z\kern-0.2em Z$}}}}

\def\diameter{{\ifmmode\oslash\else$\oslash$\fi}}

%

\skewchar\ninei='177 \skewchar\sixi='177
\skewchar\ninesy='60 \skewchar\sixsy='60
\hyphenchar\ninett=-1
\def\newline{\hfil\break}%
\catcode`@=11
\def\folio{\ifnum\pageno<\z@
\uppercase\expandafter{\romannumeral-\pageno}%
\else\number\pageno \fi}
\catcode`@=12 
  \mathchardef\Gamma="0100
  \mathchardef\Delta="0101
  \mathchardef\Theta="0102
  \mathchardef\Lambda="0103
  \mathchardef\Xi="0104
  \mathchardef\Pi="0105
  \mathchardef\Sigma="0106
  \mathchardef\Upsilon="0107
  \mathchardef\Phi="0108
  \mathchardef\Psi="0109
  \mathchardef\Omega="010A
\def\squareforqed{\hbox{\rlap{$\sqcap$}$\sqcup$}}
\def\qed{\ifmmode\squareforqed\else{\unskip\nobreak\hfil
\penalty50\hskip1em\null\nobreak\hfil\squareforqed
\parfillskip=0pt\finalhyphendemerits=0\endgraf}\fi}
\newfam\sansfam
\textfont\sansfam=\tensans\scriptfont\sansfam=\sevensans
\scriptscriptfont\sansfam=\fivesans
\def\sans{\fam\sansfam\tensans}
\def\stackfigbox{\if
Y\FIG\global\setbox\figbox=\vbox{\unvbox\figbox\box1}%
\else\global\setbox\figbox=\vbox{\box1}\global\let\FIG=Y\fi}
\def\placefigure{\dimen0=\ht1\advance\dimen0by\dp1
\advance\dimen0by5\baselineskip
\advance\dimen0by0.4true cm
\ifdim\dimen0>\vsize\pageinsert\box1\vfill\endinsert
\else
\if Y\FIG\stackfigbox\else
\dimen0=\pagetotal\ifdim\dimen0<\pagegoal
\advance\dimen0by\ht1\advance\dimen0by\dp1\advance\dimen0by1.7true cm
\ifdim\dimen0>\pagegoal\stackfigbox
\else\box1\vskip7true mm\fi
\else\box1\vskip7true mm\fi\fi\fi\let\firstleg=Y}
%
\def\begfig#1cm#2\endfig{\par
\setbox1=\vbox{\dimen0=#1true cm\advance\dimen0
by1true cm\kern\dimen0\vskip-.8333\baselineskip#2}\placefigure}
\def\begdoublefig#1cm #2 #3 \enddoublefig{\begfig#1cm%
\line{\vtop{\hsize=0.46\hsize#2}\hfill
\vtop{\hsize=0.46\hsize#3}}\endfig}
\let\firstleg=Y
\def\figure#1#2{\if Y\firstleg\vskip1true cm\else\vskip1.7true mm\fi
\let\firstleg=N\noindent{\petit{\bf
Fig.\ts#1\unskip.\ }\ignorespaces #2\smallskip}}

\catcode`@=11
\def\begtab#1cm#2\endtab{\par
\ifvoid\topins           
   \midinsert
\else
   \topinsert
\fi                      
\vbox{\vskip\topskip\hrule height\z@
         \nobreak\vskip -\topskip\vskip-9pt
         \petit#2
         \kern#1true cm\relax
         \vskip\medskipamount}%
\endinsert}
\def\begtabempty#1cm#2\endtabempty{\begtab#1cm#2\endtab}
%
\def\begtabfull#1\endtabfull{\par
\ifvoid\topins
   \midinsert 
\else
   \topinsert
\fi           
\vbox{\vskip\topskip\hrule height\z@
      \nobreak\vskip -\topskip\vskip-9pt
      \petit#1
      \vskip\medskipamount}\endinsert}
\catcode`@=12 
\def\begpet{\vskip6pt\bgroup\petit}
\def\endpet{\vskip6pt\egroup}
\newdimen\refindent
\newlinechar=`\|
\def\begref#1#2{\titlea{}{#1}%
\bgroup\petit
\setbox0=\hbox{#2\enspace}\refindent=\wd0\relax
\if!#2!\else
\ifdim\refindent>0.5em\else
\message{|Something may be wrong with your references;}%
\message{probably you missed the second argument of \string\begref.}%
\fi\fi}
\def\ref{\goodbreak
\hangindent\oldparindent\hangafter=1
\noindent\ignorespaces}
\def\refno#1{\goodbreak
\setbox0=\hbox{#1\enspace}\ifdim\refindent<\wd0\relax
\message{|Your reference `#1' is wider than you pretended in using
\string\begref.}\fi
\hangindent\refindent\hangafter=1
\noindent\kern\refindent\llap{#1\enspace}\ignorespaces}
\def\refmark#1{\goodbreak
\setbox0=\hbox{#1\enspace}\ifdim\refindent<\wd0\relax
\message{|Your reference `#1' is wider than you pretended in using
\string\begref.}\fi
\hangindent\refindent\hangafter=1
\noindent\hbox to\refindent{#1\hss}\ignorespaces}
\def\endref{\goodbreak\endpet}
\def\vec#1{{\textfont1=\tams\scriptfont1=\tamss
\textfont0=\tams\scriptfont0=\tamss
\mathchoice{\hbox{$\displaystyle#1$}}{\hbox{$\textstyle#1$}}
{\hbox{$\scriptstyle#1$}}{\hbox{$\scriptscriptstyle#1$}}}}
\def\petit{\def\rm{\fam0\ninerm}%
\textfont0=\ninerm \scriptfont0=\sixrm \scriptscriptfont0=\fiverm
 \textfont1=\ninei \scriptfont1=\sixi \scriptscriptfont1=\fivei
 \textfont2=\ninesy \scriptfont2=\sixsy \scriptscriptfont2=\fivesy
 \def\it{\fam\itfam\nineit}%
 \textfont\itfam=\nineit
 \def\sl{\fam\slfam\ninesl}%
 \textfont\slfam=\ninesl
 \def\bf{\fam\bffam\ninebf}%
 \textfont\bffam=\ninebf \scriptfont\bffam=\sixbf
 \scriptscriptfont\bffam=\fivebf
 \def\sans{\fam\sansfam\ninesans}%
 \textfont\sansfam=\ninesans \scriptfont\sansfam=\sixsans
 \scriptscriptfont\sansfam=\fivesans
 \def\tt{\fam\ttfam\ninett}%
 \textfont\ttfam=\ninett
 \normalbaselineskip=11pt
 \setbox\strutbox=\hbox{\vrule height7pt depth2pt width0pt}%
 \normalbaselines\rm
\def\vec##1{{\textfont1=\tbms\scriptfont1=\tbmss
\textfont0=\ninebf\scriptfont0=\sixbf
\mathchoice{\hbox{$\displaystyle##1$}}{\hbox{$\textstyle##1$}}
{\hbox{$\scriptstyle##1$}}{\hbox{$\scriptscriptstyle##1$}}}}}
\nopagenumbers
%
\let\header=Y
\let\FIG=N
\newbox\figbox
\output={\if N\header\headline={\hfil}\fi\plainoutput
\global\let\header=Y\if Y\FIG\topinsert\unvbox\figbox\endinsert
\global\let\FIG=N\fi}
\let\lasttitle=N
\catcode`\@=\active
\def\author#1{\bgroup
\parindent=0pt
\baselineskip=13.2pt
\lineskip=0pt
\pretolerance=10000
\rm
\ignorespaces#1\par\bigskip\egroup
{\def@##1{}%
\setbox0=\hbox{\petit\kern2.5true cc\ignorespaces#1\unskip}%
\ifdim\wd0>\hsize
\message{The names of the authors exceed the headline, please use a }%
\message{short form with AUTHORRUNNING}\gdef\leftheadline{%
\rlap{\folio}\hfil AUTHORS suppressed due to excessive length}%
\else
\xdef\leftheadline{\rlap{\noexpand\folio}
\kern15pt#1\hfil}
\fi}
\let\INS=E}
\def\address#1{\bgroup\petit
\parindent=0pt
\ignorespaces#1\par\bigskip\egroup
\catcode`\@=12
\noindent\ignorespaces}
\let\INS=N%

\def@#1{\if N\INS\unskip$^{#1}$%
\else\global\footcount=#1\relax
\if E\INS\noindent$^{#1}$\enspace\let\INS=Y\ignorespaces
\else\par\noindent$^{#1}$\enspace\ignorespaces\fi\fi}%
\catcode`\@=12
\headline={\petit\def\newline{ }\def\fonote#1{}\ifodd\pageno
\rightheadline\else\leftheadline\fi}
\def\rightheadline{Missing CONTRIBUTION
title\hfil\llap{\folio}}
\def\leftheadline{\rlap{\folio}\hfil Missing name(s)
of the author(s)}
\nopagenumbers
\let\header=Y

\let\lasttitle=N
 \def\contribution#1{\vfill\eject
 \let\header=N\bgroup
 \parindent=0pt
 \textfont0=\tafontt \scriptfont0=\tafonts

 \scriptscriptfont0=\tafontss
 \textfont1=\tamt \scriptfont1=\tams

 \scriptscriptfont1=\tams
 \textfont2=\tast \scriptfont2=\tass

 \scriptscriptfont2=\tasss
 \par\baselineskip=16pt
     \lineskip=16pt
     \tafontt
     \pretolerance=10000
     \noindent
     \ignorespaces#1\par
   \vskip4.2cm\egroup
    \nobreak
     \parindent=0pt
     \everypar={\global\parindent=1.5em
     \global\let\lasttitle=N\global\everypar={}}%
     \global\let\lasttitle=A%
      \setbox0=\hbox{\petit\def\newline{ }
      \def\fonote##1{}\kern2.5true
      cc\ignorespaces#1}\ifdim\wd0>\hsize
      \message{Your CONTRIBUTIONtitle exceeds the headline,
 please use a short form
 with CONTRIBUTIONRUNNING}\gdef\rightheadline{CONTRIBUTION

 title suppressed due to excessive length\hfil\llap{\folio}}%
 \else
\gdef\rightheadline{\hfil\ignorespaces#1\kern15pt\llap{\folio}}\fi
\catcode`\@=\active
     \ignorespaces}
\def\titlea#1#2{\if N\lasttitle\else\vskip-28pt
     \fi
     \vskip18pt plus 4pt minus4pt
     \bgroup
\textfont0=\tbfontt \scriptfont0=\tbfonts

\scriptscriptfont0=\tbfontss
\textfont1=\tbmt \scriptfont1=\tbms

\scriptscriptfont1=\tbmss
\textfont2=\tbst \scriptfont2=\tbss
 \scriptscriptfont2=\tbsss
\textfont3=\tasys \scriptfont3=\tenex

\scriptscriptfont3=\tenex
     \baselineskip=16pt
     \lineskip=0pt
     \pretolerance=10000
     \noindent
     \tbfontt
     \rightskip 0pt plus 6em
     \setbox0=\vbox{\vskip23pt\def\fonote##1{}%
     \noindent
     \if!#1!\ignorespaces#2
     \else\ignorespaces#1\unskip\quad\ignorespaces#2\fi
     \vskip18pt}%
     \dimen0=\pagetotal\advance\dimen0 by-\pageshrink
     \ifdim\dimen0<\pagegoal
     \dimen0=\ht0\advance\dimen0 by\dp0\advance\dimen0 by
     3\normalbaselineskip
     \advance\dimen0 by\pagetotal
     \ifdim\dimen0>\pagegoal\eject\fi\fi
     \noindent
     \if!#1!\ignorespaces#2
     \else\ignorespaces#1\unskip\quad\ignorespaces#2\fi
     \vskip12pt plus4pt minus4pt\egroup
     \nobreak
     \parindent=0pt
     \everypar={\global\parindent=\oldparindent
     \global\let\lasttitle=N\global\everypar={}}%
     \global\let\lasttitle=A%
     \ignorespaces}
 \def\titleb#1#2{\if N\lasttitle\else\vskip-22pt
     \fi
     \vskip18pt plus 4pt minus4pt
     \bgroup
\textfont0=\tenbf \scriptfont0=\sevenbf

\scriptscriptfont0=\fivebf
\textfont1=\tams \scriptfont1=\tamss

\scriptscriptfont1=\tbmss
     \lineskip=0pt
     \pretolerance=10000
     \noindent
     \bf
     \rightskip 0pt plus 6em
     \setbox0=\vbox{\vskip23pt\def\fonote##1{}%
     \noindent
     \if!#1!\ignorespaces#2
     \else\ignorespaces#1\unskip\quad\ignorespaces#2\fi
     \vskip10pt}%
     \dimen0=\pagetotal\advance\dimen0 by-\pageshrink
     \ifdim\dimen0<\pagegoal
     \dimen0=\ht0\advance\dimen0 by\dp0\advance\dimen0 by
     3\normalbaselineskip
     \advance\dimen0 by\pagetotal
     \ifdim\dimen0>\pagegoal\eject\fi\fi
     \noindent
     \if!#1!\ignorespaces#2
     \else\ignorespaces#1\unskip\quad\ignorespaces#2\fi
     \vskip8pt plus4pt minus4pt\egroup
     \nobreak
     \parindent=0pt
     \everypar={\global\parindent=\oldparindent
     \global\let\lasttitle=N\global\everypar={}}%
     \global\let\lasttitle=B%
     \ignorespaces}
 \def\titlec#1{\if N\lasttitle\else\vskip-\baselineskip
     \fi
     \vskip12pt plus 4pt minus4pt
     \bgroup
\textfont0=\tenbf \scriptfont0=\sevenbf
 \scriptscriptfont0=\fivebf
\textfont1=\tams \scriptfont1=\tamss

\scriptscriptfont1=\tbmss
     \bf
     \noindent
     \ignorespaces#1\unskip\ \egroup
     \ignorespaces}
 \def\titled#1{\if N\lasttitle\else\vskip-\baselineskip
     \fi
     \vskip12pt plus 4pt minus 4pt
     \bgroup
     \it
     \noindent
     \ignorespaces#1\unskip\ \egroup
     \ignorespaces}
\let\ts=\thinspace
\def\footnoterule{\kern-3pt\hrule width 2true cm%
\kern2.6pt}
\newcount\footcount \footcount=0
\def\advftncnt{\advance\footcount by1%
\global\footcount=\footcount}
\def\fonote#1{\advftncnt$^{\the\footcount}$%
\begingroup\petit
\parfillskip=0pt plus 1fil
\def\textindent##1{\hangindent0.5\oldparindent%
\noindent\hbox to0.5\oldparindent{##1\hss}%
\ignorespaces}%
\vfootnote{$^{\the\footcount}$}{#1\vskip-9.69pt}%
\endgroup}
\def\item#1{\par\noindent
\hangindent6.5 mm\hangafter=0
\llap{#1\enspace}\ignorespaces}

\def\newenvironment#1#2#3#4{\long\def#1##1##2%
{\removelastskip
\vskip\baselineskip\noindent{#3#2\if!##1!.\else\unskip\ %
\ignorespaces
##1\unskip\fi\ }{#4\ignorespaces##2}\vskip\baselineskip}}
\newenvironment\lemma{Lemma}{\bf}{\it}
\newenvironment\proposition{Proposition}{\bf}{\it}
\newenvironment\theorem{Theorem}{\bf}{\it}
\newenvironment\corollary{Corollary}{\bf}{\it}
\newenvironment\example{Example}{\it}{\rm}
\newenvironment\exercise{Exercise}{\bf}{\rm}
\newenvironment\problem{Problem}{\bf}{\rm}
\newenvironment\solution{Solution}{\bf}{\rm}
\newenvironment\definition{Definition}{\bf}{\rm}
\newenvironment\note{Note}{\it}{\rm}
\newenvironment\question{Question}{\it}{\rm}
\long\def\remark#1{\removelastskip\vskip\baselineskip%
\noindent{\it
Remark.\ }\ignorespaces}
\long\def\proof#1{\removelastskip\vskip\baselineskip%
\noindent{\it
Proof\if!#1!\else\ \ignorespaces#1\fi.\ }\ignorespaces}
\def\abstract#1{\par\vskip3mm{\petit
\noindent{\bf Abstract. }\ignorespaces#1\par}\vskip1mm}
\def\typeset{\petit\noindent This article was processed

by the author
using the plain\TeX\ PESO macro package from

Springer-Verlag.\par}
\outer\def\byebye{\bigskip\bigskip\typeset
\footcount=1\ifx\speciali\undefined\else
\loop\smallskip\noindent special character No\number%
\footcount:
\csname special\romannumeral\footcount\endcsname
\advance\footcount by 1\global\footcount=\footcount
\ifnum\footcount<11\repeat\fi
\vfill\supereject\end}

\contribution{Colour Gradients in the Optical and Near-IR}
\author{Roelof S.~de Jong@1@2}
\address{
@1Univ. of Durham, Dept. of Physics, South Road, Durham DH1 3LE, UK
@2Kapteyn Institute, P.O.box 800, 9700 AV Groningen, The Netherlands
}
\abstract{
 For many years broadband colours have been used to obtain insight into
the contents of galaxies, in particular to estimate stellar and dust
content.  Broadband colours are easy to obtain for large samples of
objects, making them ideal for statistical studies.  In this paper I use
the radial distribution of the colours in galaxies, which gives more
insight into the local processes driving the global colour differences
than integrated colours.  Almost all galaxies in my sample of 86 face-on
galaxies become systematically bluer with increasing radius.  The radial
photometry is compared to new dust extinction models and stellar
population synthesis models.  This comparison shows that the colour
gradients in face-on galaxies are best explained by age and metallicity
gradients in the stellar populations and that dust reddening plays
a minor role.  The colour gradients imply $M/L$ gradients, making the
`missing light' problem as derived from rotation curve fitting even
worse.
}

\titlea{1}{The colour gradients}
 A sample of 86 spiral galaxies was imaged in the $B$, $V$, $R$, $I$,
$H$ and $K$ passbands to study light and colour distributions as a
function of radius.  Full details of sample selection and data reduction
are described in de Jong \& van der Kruit (1994).  The galaxies were
selected to be face-on and to have a diameter of at least 2\arcmin.  The
sample is statistically complete and can be corrected for selection
effects.  It can therefore be used to analyze the nature of the Freeman
law (Freeman 1970) and this analysis has been reported elsewhere (de
Jong 1995a, 1995b).

The luminosity profiles were determined in the usual way by measuring
the average surface brightness on annuli of increasing radius.  Radial
colour profiles were created by combining profiles in different
passbands.  The run of colour as function of radius is put on a common
scale for all galaxies in Fig.~1, where the average $B$--$K$ colour at
each radius is plotted as function of the average $R$ surface brightness
at this radius.

\begfig 0 cm
 \epsfysize=5.9cm \epsfbox[55 480 560 750]{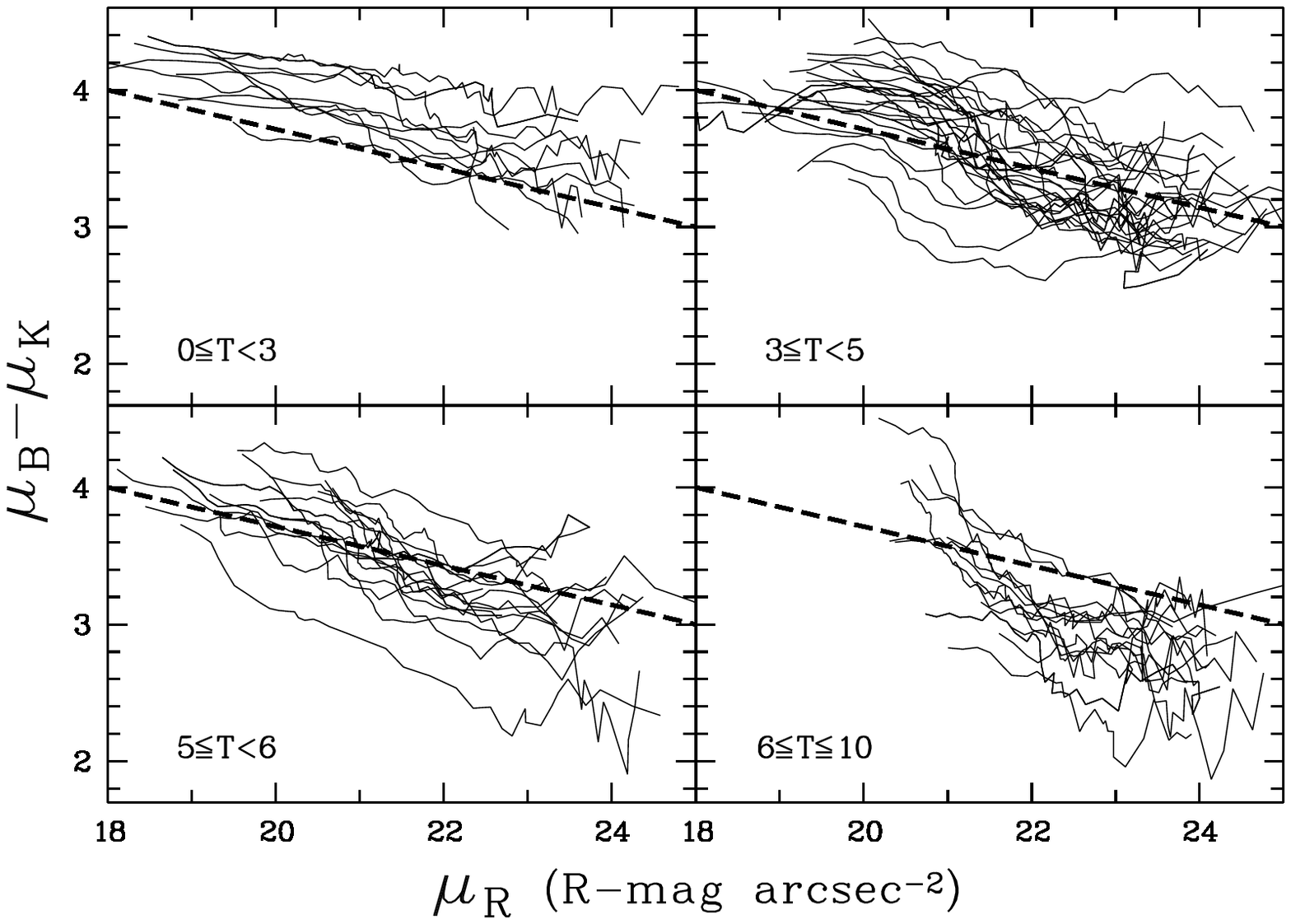}
 \figure{1}{The average $B$--$K$ colour at each radius as function of
the average $R$ surface brightness at this radius.  The galaxies are
divided into 4 morphological RC3 T-type bins. The dashed lines are
provided to have a common reference among the bins.}
 \endfig

Two observations can readily be made from this diagram.  Firstly, all
galaxies become bluer going radially outward, correlating strongly with
the average surface brightness at each radius.  Secondly, even at the
same surface brightness, late type spiral galaxies are bluer than
earlier types (use the dashed lines to guide the eye).  Furthermore it
should be noted that there is a smooth transition in colour from the
bulge to the disk region.  The colours of bulges are nearly identical to
the colours of inner disk regions (see also Peletier, these
proceedings).

\titlea{2}{The dust and stellar population synthesis models}
 A possible explanation for the colour gradients is reddening due to dust
extinction.  As galaxies are intricate mixtures of stars and dust, we
cannot describe the reddening by a simple extinction law, but have to
calculate the separate contributions of absorption and scattering.  To
predict reddening profiles due to absorption and scattering full 3D
Monte Carlo simulations were made (de Jong 1995a) of galaxies with
smooth exponential dust and stellar distributions in both radial and
vertical directions.  The main free parameters are the dust to stellar
scaleheight ($z_{d}/z_{s}$) and scalelength ($h_{d}/h_{s}$) ratios, the
central optical depth ($\tau_{0,V}$) and the properties of the dust
particles.  The (poorly determined) Galactic dust properties were
assumed, since extra-galactic albedo and phase scattering functions have
never been measured.

\begfig 0 cm
\hbox{
\epsfysize=4.7cm
\epsfbox[18 231 560 670]{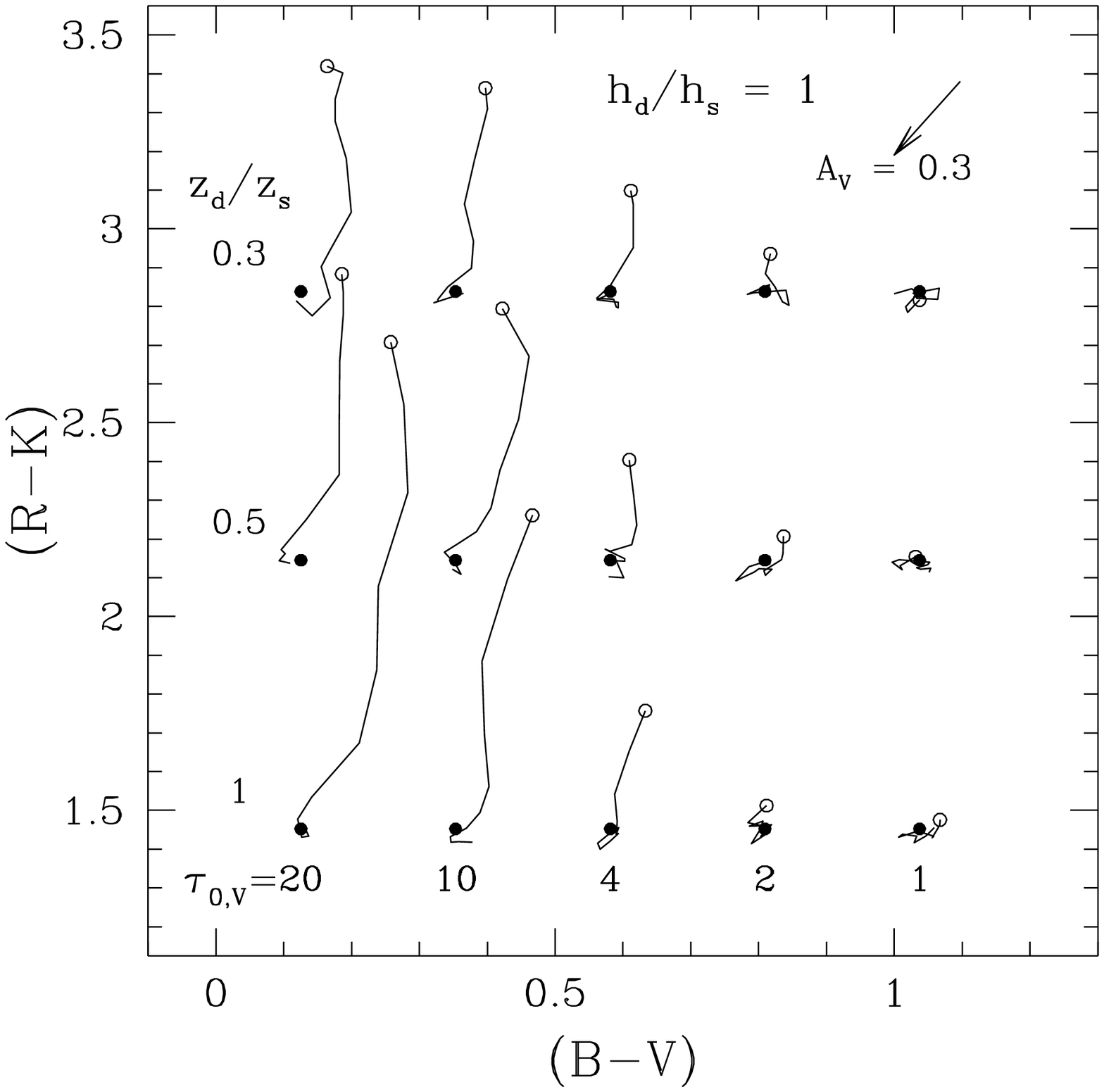}
\epsfysize=4.7cm
\epsfbox[18 231 560 670]{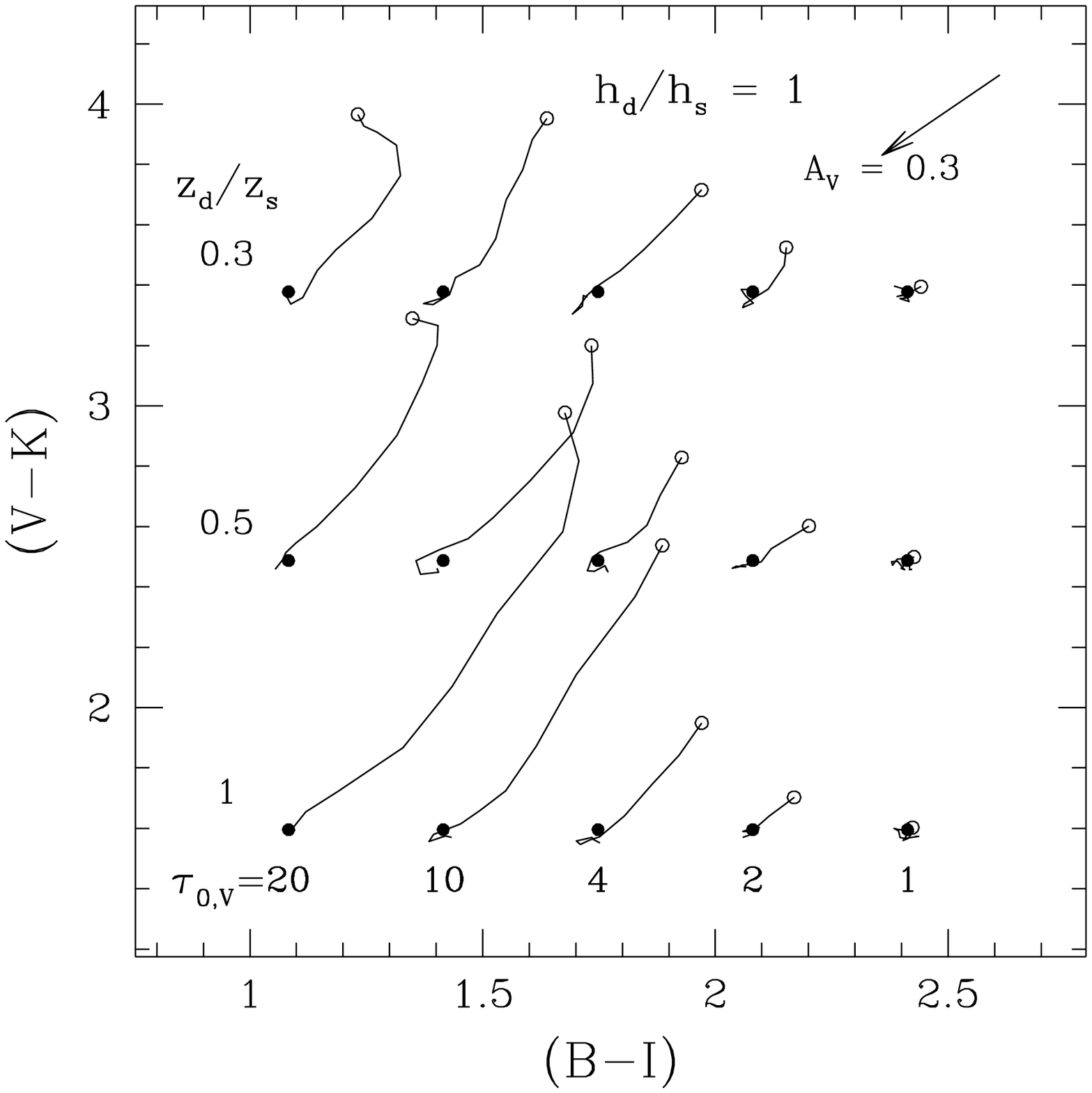}
}
 \figure{2}{The radial colour-colour reddening profiles resulting from
the Monte Carlo simulations of stars and dust in exponential disks for
different $z_{d}/z_{s}$ and $\tau_{0,V}$ values.  The arbitrary
unreddened colours are indicated with filled circles for each model, the
galaxy centres indicated by open circles.
}
 \endfig

Figure 2 shows a number of model colour-colour reddening profiles for
different $z_{d}/z_{s}$ and $\tau_{0,V}$ values.  The positions of the
models in the diagram are arbitrary (depending on the colours of the
underlying population), but the shape of the reddening profiles are
determined by the distribution and the properties of the dust.  Note
that all reddening profiles point in about the same direction, independent
of the dust configuration, and that this direction is different from the
standard screen model extinction vector (arrow).

 I have used two sets of stellar population synthesis models in the
comparison with the data.  The Solar metallicity models of Bruzual \&
Charlot (1993) are used to study the colour changes of
populations due to different star formation histories.  Two extreme
cases are considered, a single star burst model and a constant star
formation model.  The colours of these populations are inspected after 8
and 17$\,$Gyr.  The models of Worthey (1994) are used to study the effects
that age and metallicity have on the colours of a population.

\titlea{3}{The comparison between models and data}

\begfig 0 cm
 \epsfysize=7cm \epsfbox[18 340 545 655]{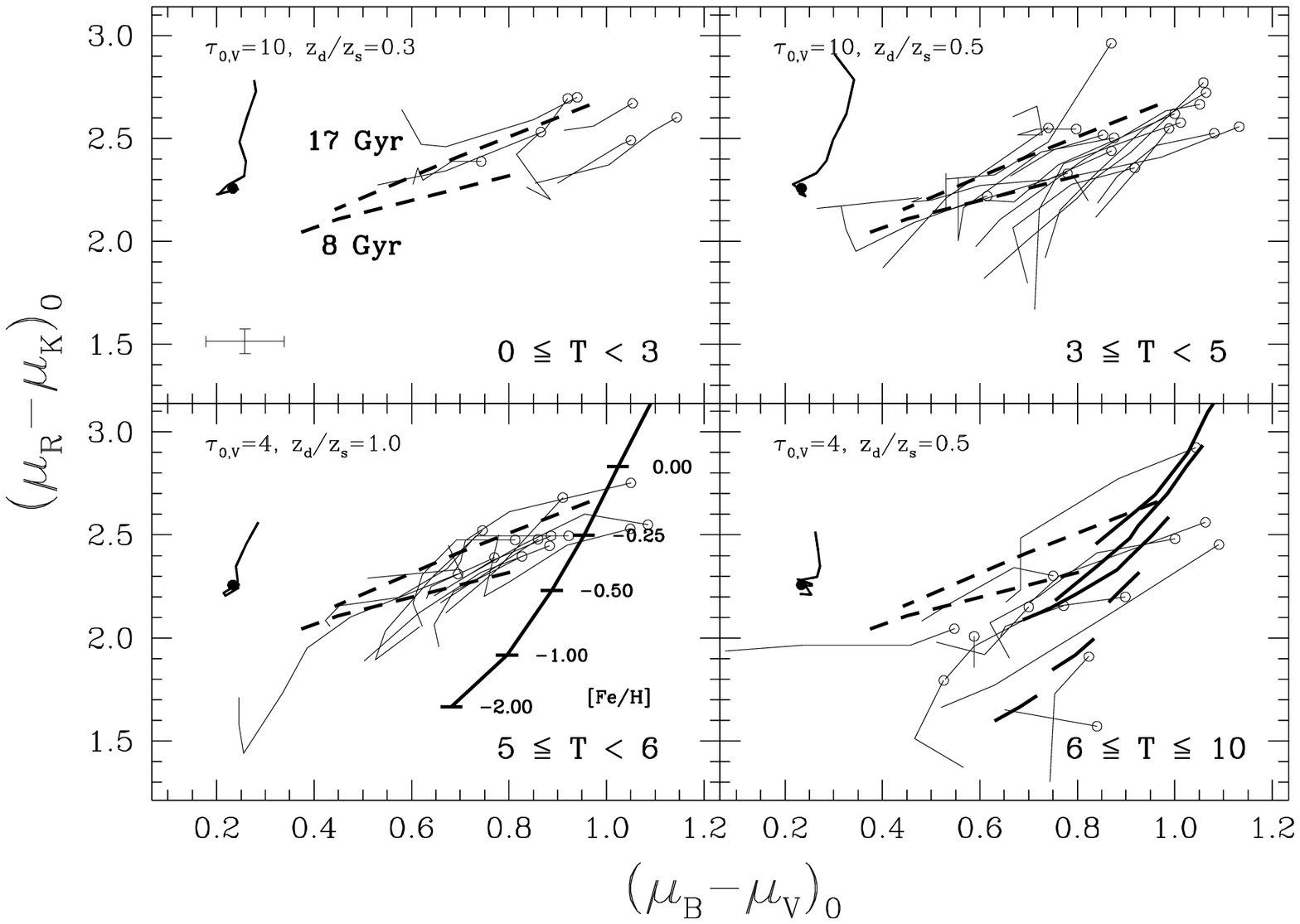}
 \figure{3a}{
 The run of $B$--$V$ versus $R$--$K$ colour as function of radius for
the galaxies (thin lines, centre indicated by open circle).  Dust models
in the top left corner, Bruzual \& Charlot (1993) models after 8 and 17$\,$Gyr
thick dashed lines, Worthey (1994) models for indicated
metallicities thick solid lines in the two bottom panels.
}
 \endfig

 Figure 3 shows the colour-colour diagrams of the models and the
galaxies, again divided into four T-type bins.  Note that the {\it same}
model should fit the data in {\it all} colour combinations, thus in
both Figs.  3a \& 3b.  The thin lines represent the galaxy data; the
central galaxy colours are indicated by the open circles, the lines show
the run of colours as function of radius.
 All galaxies with type T$<$6 are confined to a small region in these
diagrams, only the later types show a considerable larger spread.

Some dust model profiles are indicated in the top-left corner of the
panels.  As mentioned in Sect.~2, these profiles can be placed anywhere
in the diagrams and their direction depends mainly on the dust
properties, not on the relative distribution of dust and stars.
Clearly, the colour gradients cannot be caused by reddening alone,
assuming that the dust properties used are correct.

\begfig 0 cm
\epsfysize=7cm
\epsfbox[18 340 545 655]{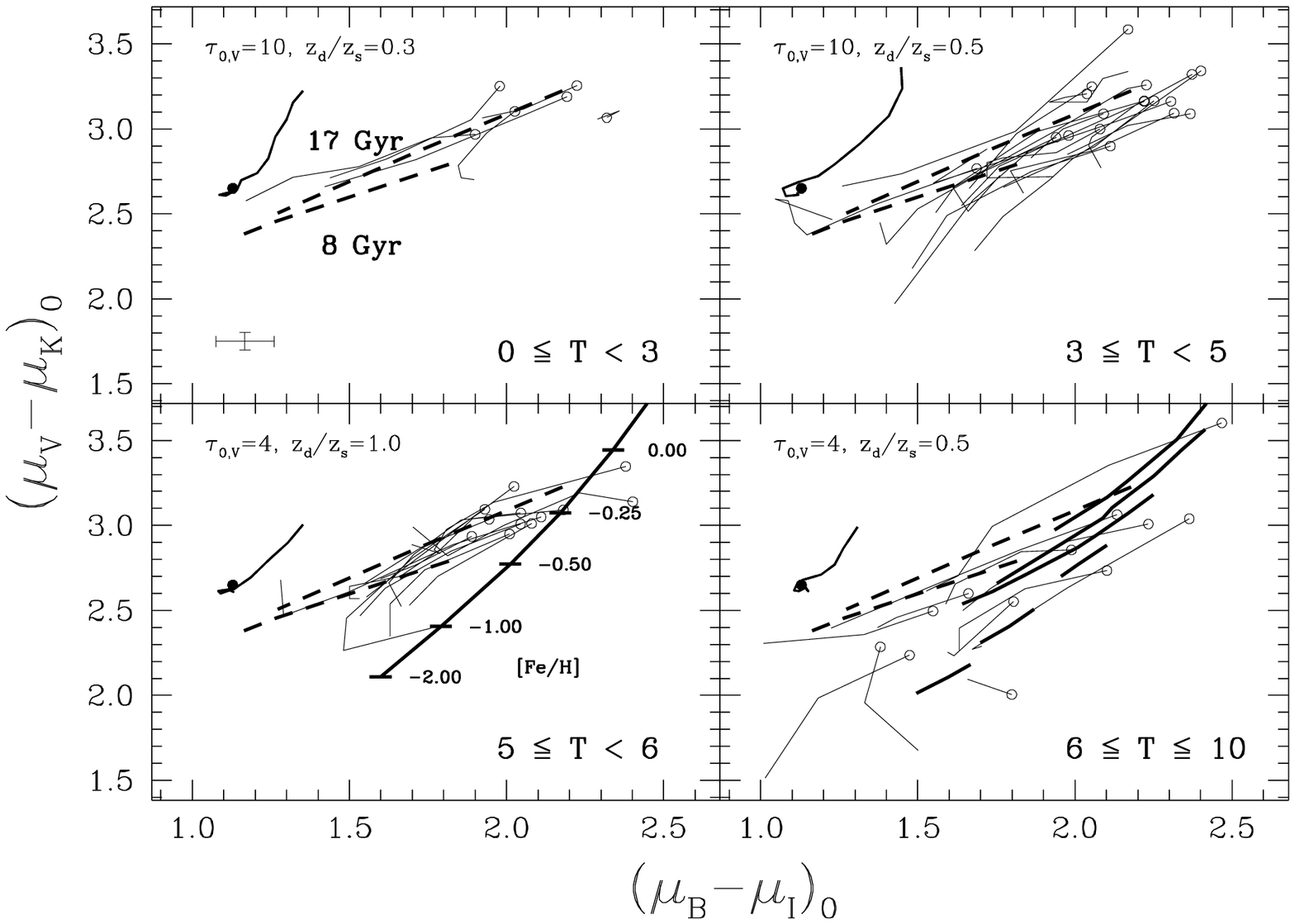}
\figure{3b}{
As Fig.~3a, but for $B$--$I$ versus $V$--$K$.
}
\endfig

In the 5$\le$T$<$6 panels the colours predicted by Worthey's models
after 12$\,$Gyr are shown for a range of indicated metallicities. The
metallicity-colour trend runs in the same direction for other ages and
apparently, a metallicity gradient alone cannot explain the observed
colour gradients.

The effects of different star formation histories are indicated by the
two dashed lines in the centre of the panels.  The red ends of these
lines indicate the colours of a single burst population, the blue end of
a population of constant star formation rate.  Both the position and the
direction of these solar metallicity models seem to agree reasonably
well with the data, and the most simple explanation for the colour
gradients would be age gradients across the disks of spiral galaxies.
Still this cannot be the whole story, as galaxies still have star
formation in their central regions, which means that a single burst is a
bad approximation.  Furthermore it is known from measurement of HII
regions that spiral galaxies have metallicity gradients in their gas
content, which most likely is partly reflected in the stellar component.
So the most consistent picture is one where colour gradients are
caused by both age and metallicity gradients, with the central regions
of galaxies being on average quite old and having a range of
metallicities, whereas the outer parts are young and have low
metallicities.

The large spread in colours of the late type galaxies can be explained
by stellar population changes as well.  The single-burst age evolution
is indicated by the thick, solid lines in the 6$\le$T$\le$10 panels,
for the metallicities indicated in the 5$\le$T$<$6 panels.  The colours
indicate that the stellar population in some of these galaxies are on
average very young and have a low metallicity.

The colour gradients imply large $M/L$ gradients for the optical
passbands, making the `missing light' problem as derived from rotation
curve fitting even worse, irrespective whether they are caused by dust
or population changes.

\begref{References}{ }
 \ref Bruzual G.A., Charlot S. 1993, ApJ 405, 538
 \ref de Jong R.S. 1995a, Ph.D.~Thesis, Univ. of Groningen, The
Netherlands
 \ref de Jong R.S. 1995b, A\&A, submitted
 \ref de Jong R.S., van der Kruit P.C. 1994, A\&AS 106, 451
 \ref Freeman K.C. 1970, ApJ 160, 811
 \ref Worthey G. 1994, ApJS 95, 107

\endref

\bye